# SMART: Investigating the Impact of Threshold Voltage Suppression in an In-SRAM Multiplication/Accumulation Accelerator for Accuracy Improvement in 65 nm CMOS Technology

Saeed Seyedfaraji*, Baset Mesgari*, Semeen Rehman*
*Vienna University of Technology (TU-Wien), Vienna, Austria
{saeed.seyedfaraji; baset.mesgari, semeen.rehman}@tuwien.ac.at

*Abstract*—State-of-the-art in-memory computation has recently emerged as the most promising solution to overcome design challenges related to data movement inside current computing systems. One of the approaches to performing in-memory computation is based on the analog behavior of the data stored inside the memory cell. These approaches proposed various system architectures for that. In this paper, we investigated the effect of threshold voltage suppression on the access transistors of the In-SRAM multiplication and accumulation (MAC) accelerator to improve and enhance the performance of bit line (bit line bar) discharge rate that will increase the accuracy of MAC operation. We provide a comprehensive analytical analysis followed by circuit implementation, including a Monte-Carlo simulation by a 65nm CMOS technology. We confirmed the efficiency of our method (SMART) for a four-by-four-bit MAC operation. The proposed technique improves the accuracy while consuming 0.683 pJ per computation from a power supply of 1V. Our novel technique presents less than 0.009 standard deviations for the worst-case incorrect output scenario.

## I. INTRODUCTION

For many years Von-Neumann architecture has been widely used for system designing. It includes a processing unit (processing element and caches), the central memory unit, and I/O devices that have been connected through the system Bus [1], [2]. For performing a logical/mathematical operation, the processor needs to fetch the data through the specific protocols and store the operands inside the register files. This data movement will result in energy and latency overhead on the system [3]. This concept, known as memory-wall, stands as the most critical barrier to energy-efficient system designs [4], [5]. A recent study revealed that more than 62.7% of total system energy is used on data movement in everyday applications such as google chrome, Tensor Flow Mobile, and Video Playback. Since data-intensive applications (e.g., Neural Networks, Machine Learning, Deep Learning, etc.) are fundamentally important nowadays, the energy consumption caused by data movement becomes vital in system design, specifically for resource constraint embedded applications [3], [6], [7], [8].

The most effective solution for overcoming this memory barrier is realized through the concept of in-memory processing (IMP). The IMP refers to equipping the memory elements (CACHEs, main memory) with substantial processing capabilities, allowing the system to perform basic mathematical or logical operations inside the memory element without moving the data [9]–[12]. Considering the fact that the applications mentioned above mostly perform a unique mathematical function repeatedly i.e., Multiplication and Accumulation (MAC), IMP delivers a solution to accomplish this operation inside the memory element and eliminates the data movement, providing energy saving. Different memory technologies (including volatile and non-volatile memories) can perform IMP. There has been a broad debate over each technology's advantages and disadvantages [9], [10], [13], [14], [15], [16]. We will present our novel IMP circuit over Static Random-Access Memory (SRAM) due to two main features. First, this memory technology is highly divisible, which allows the end-user to access different memory banks simultaneously (parallelism), and has a low fabrication cost leading the SRAM as one of the most common industrial on-chip technology [9], [10], [13], [17]. The current state-of-the-art in-SRAM techniques utilize the "read" operation of the standard memory to perform the mathematical function (see Fig. 1). First, one operand (e.g., operand "A" in the case of A*B, or A+B) will be written to the memory cell, and then the other operand "B" will be passed through a Digital to Analog Converter (DAC) to be codded to an appropriate voltage. This voltage then activates the access transistors of SRAM. A charge sharing on the bit-line (BL) or bit-line-bar (BLB) will be conducted and finally interpreted as the operation's result [9], [10], [14], [18], [19] (see Fig. 1).

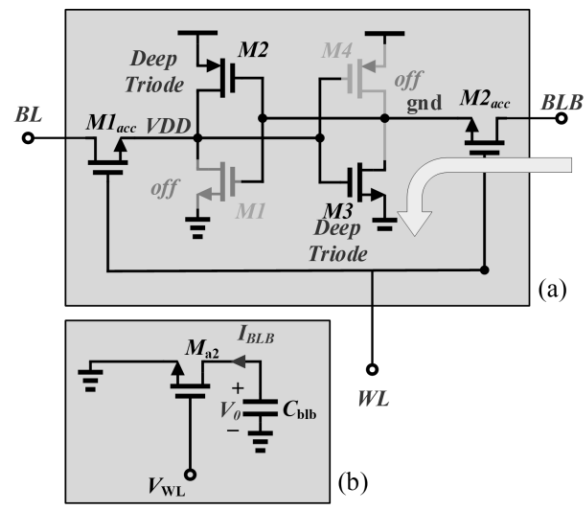

Fig. 1. (a) Standard single bit 6T-SRAM memory cell in the read mode and (b) its' discharge equivalent circuit.

Once WL is enabled, the only possibility of establishing a substantial current flow is through the transistors named $M2_{acc}$ and $M_3$. Consequently, $M2_{acc}$ creates a path for BLB to discharge the initial charge of $C_{blb}$ from VDD towards the ground (0V). This discharge voltage should be sampled and appropriately combined with other samples collected from other memory cells. This combination later will be exploited to an accurate interpretation output based on the analog voltage on WL and digital stored code inside each bit (SRAM-cell). Therefore the dynamic behavior of current through the $M2_{acc}$ defines the accuracy of analog multiplication/sum. The available voltage range for the WL determines the accuracy of the output current and the maximum number of coded bits, which is usually limited to the transistor's threshold voltage ($V_{TH}$). In [9], [10], 350 mV margin has been chosen to prevent nonlinear behavior of BLB voltage drop for $M2_{acc}$ transistor implemented in 65nm CMOS technology. Based on what has been presented in [9] (and will be elaborated in section 2), if the power supply is equal to VDD, while $V_{TH}$ is driven near zero, the accuracy in generating the coded analog data could be improved by ($V_{TH}$ / (VDD-$V_{TH}$) *100) percent in the ideal case. Additionally, by controlling the voltage of the bulk pin of M2acc (Fig.1.a), the body effect of this transistor can be suppressed. Of course, for providing this method, you need to use a deep-nwell transistor for M2acc, which is convenient in a standard CMOS process. In this paper, to increase the WL maximum pulse width ($WL_{PW\text{-}MAX}$), we propose a novel circuit designed based on the body biasing technique to suppress $V_{TH}$, leading to an increased voltage range for WL. To the best of the authors' knowledge, SMART is the first manuscript that investigates the effect of VTH suppression in the in-memory MAC and enhances the accuracy of the results. The rest of the paper is organized as follows: Section II provides an overview of the SRAM operation to understand how various circuit level parameters are selected in order to perform an analog multiplication, and the accuracy of the proposed structure is presented. Section III provides our novel multi-bit multiplication design, followed by section VI presenting the results and experimental setup. The SMART circuit-level simulation, the process variation analysis, and state-of-the-art comparisons are provided. We conclude this manuscript in section V.

## II. ANALOG IN-SRAM MAC OPERATION

Fig. 2 illustrates the fundamental building blocks of a 6T-SRAM memory structure. The two fundamental operations on typical memory elements are "Read" and "Write". The procedure for performing read operations is as follows: First, the preprocessing setup is required, which means that to read the cell, the WL is almost equal to zero V, and therefore assuming the Q is equal to zero, the $Q_{bar}$ will be equal to one. Then the BL and BLB should be charged up to VDD, which means the WL of the access transistors ($M1_{acc}$, $M2_{acc}$) will be connected to high voltage. As a result, the gate to source voltage ($V_{GS}$) of $M2_{acc}$ is almost equal to zero ($V_{WL} = V_{Qbar} = VDD$), which prevents channel formation, and no current will be passed (i.e., both $V_{BLB}$ and $V_{Qbar}$ remain unchanged). Since the SRAM consists of two back-to-back inverters, on the other side, the BL will be discharged gradually toward zero through M1 and M1acc, which results in a read operation of the stored value. If the initial condition is vice versa (i.e., Q = one and $Q_{bar}$ = zero), the procedure will be flipped similarly.

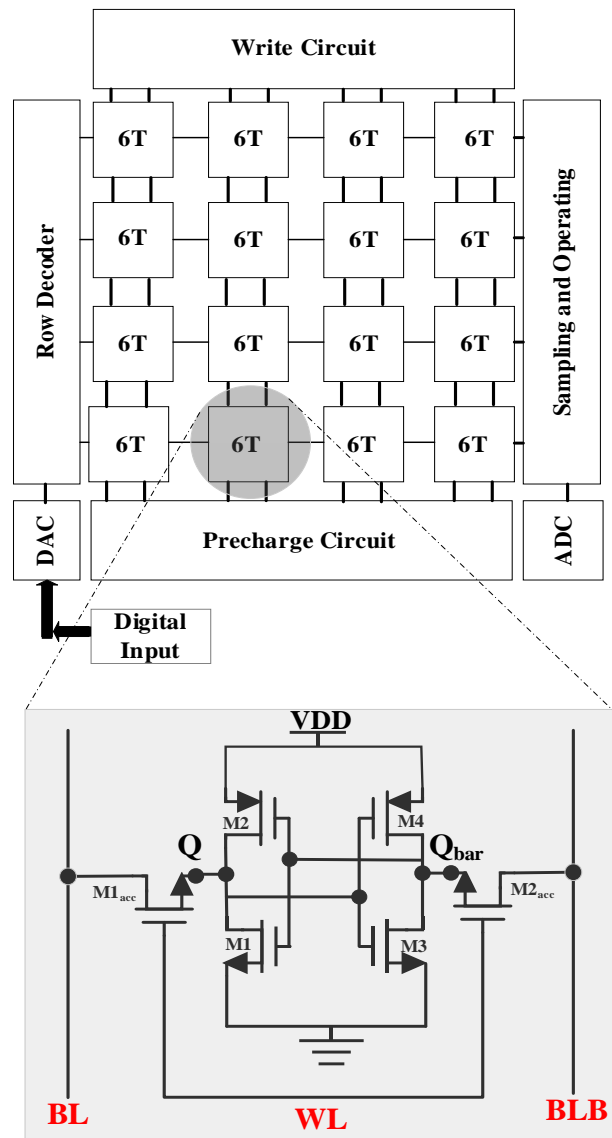

Fig. 2 SRAM structure and the 6T-SRAM cell.

However, different charging scenario of BL and BLB are required to perform a write operation. For writing logic one to the cell, the BL should be charged up to VDD while the BLB is discharged to zero, and then the WL will be connected to VDD. Writing a logic zero will be the mirror of the same procedure. In order to execute the analog MAC operation, SMART will exploit the read operation of the SRAM cell with the initial condition of Q = VDD and $Q_{bar}$ = Zero. It is noteworthy that the initial condition selected above has no positive/negative impact on the circuit design and the output results. As mentioned previously, the second digital operand of the MAC will be coded into an appropriate analog interpretation and injected into the circuit by the WL signal amplitude. We will discuss the in-SRAM analog multiplication in detail next.

## A. Quantitative Study of Analog MAC Operation Using a Standard 6T-SRAM cell

As discussed in the previous section, after performing the pre-setup conditions for a read operation, the second operand will be modulated into a voltage passing to the gate of the access transistors (WLs). The result will be calculated based on the amount of discharge absorbed by sampling capacitance of the BLB line. Before discussing the novel contribution of this paper and discuss the effect of body-biasing on the output result, first, we need to understand the effect of the voltage amplitude on the accuracy of the final result (known as bit error rate (BER)). To perform the MAC operation in the startup phase, as we did in the read operation, Q will be connected to high voltage (Q=VDD), and therefore $Q_{bar}$ will be equal to zero. Then, we charge the BL and BLB to the high voltage ($V_{BL} = V_{BLB} =$ VDD). It means that in this specific time, M3 and M2 are in their deep triode region and M2 and M3 are in the cut-off region. Once the WL has been activated, the only available way to discharge the BLB capacitance is via the path from $C_{blb}$ towards M3. Therefore, Fig. 1-b is the equivalent small signal of the primary circuit. To quantify the current ($I_{BLB}$) passing through this equivalent, we need to apply Kirchhoff's Current/Voltage Law (KVL/KCL) as Eq. 1. The current passing through the MOSFET ($I_0 = I_{BLB}$) could be calculated via Eq. 2, ignoring the channel length modulation.

$$I_{BLB} + C_{BLB} \frac{dV_{BLB}(t)}{dt} = 0 \quad (1)$$

$$I_0 = \frac{1}{2} \mu_n C_{OX} \left(\frac{W}{L}\right) \left(V_{GS_{Acc2}} - V_{TH_{Acc2}}\right)^2 \quad (2)$$

In these equations, $\mu_n$ is the mobility of the electron, and $C_{ox}$ is the gate-oxide capacitance, W/L expresses the gate's size of the MOSFET based on the length and the width, respectively. $V_{TH}$ shows the threshold voltage, and it is the minimum voltage difference between the gate and the source of the transistor ($V_{GS}$) required for the channel to be created. The channel length modulation could be applied by parameter λ in the current equation of $M2_{acc}$, and therefore equation (2) needs to be multiplied by a factor ($1+\lambda V_{BLB}$). In order to calculate the $V_{BLB}$, we combine equations 1 and 2 into one and take an integral from both sides and extract Eq. 3 based on the voltage drop of the BLB.

$$V_{BLB} = VDD - \frac{\mu_n C_{OX} \left(\frac{W}{L}\right) (V_{WL} - V_{TH})^2 t}{2 C_{BLB}} \quad (3)$$

Although the state-of-the-art approaches [9], [10] took different ways of implementing the WL, aiming to make the BLB voltage drop linearly, there are no additional parameters involved in the equation of their $V_{BLB}$ compared to what stated above. It is worth mentioning that Eq. (2) is accurate while $M2_{acc}$ remains at its saturation region. But we know that statement is for an ideal situation, and in reality, the current leaks through $M2_{acc}$ and M3 (labeled with a yellow arrow in Fig. 1) Thus, sampling the discharge of the BLB should be applied in the same region (saturation). If the access transistor falls into the triode region before the sampling time, it will result in a systematic fault in the output, and that data would be invalid. In order to confirm the accuracy of sampling, a maximum word line pulse width ($WL_{PW\_MAX}$) could be calculated based on the concern mentioned above along with the saturation condition of an NMOS transistor, and Eq. 5 could be acquired.

$$WL_{PW\_MAX} = \frac{C_{BLB}}{I_0}(V_{DD} + V_{TH} - V_{WL}) \quad (4)$$

Before discussing the $V_{TH}$ effect on the output of the MAC operation, the $V_{WL}$ should be extracted. For the case of N×N multiplication (N is the bit width of the input data and is a number in the range of $[0, 2^{N-1}]$), the WL of the circuit needs to have $2^N$ identical voltage levels to present each input data. Exploiting the KVL on the circuit, the $V_{WL}$ could be achieved as follows:

$$V_{WL} = V_{TH} + \left(V_{DAC} + \frac{(VDD - V_{TH})}{2^N - 1}\right) \quad (5)$$

Considering both Eq. (4) and (5), it is clear that increasing the $V_{TH}$ will result in a bigger pulse width available for $WL_{PW\_MAX}$ and a more significant margin for the WL itself. Therefore $V_{TH}$ is a design parameter to improve the read margin. Providing a more significant margin for the WL will improve the accuracy of the In-SRAM MAC accelerator, which will be discussed numerically in the next sections of the paper.

## B. Body Effect of a MOSFET

In most designs, including the in-SRAM MAC accelerator in Fig. 1, the designer tacitly considered that the transistor's bulk and source were tied to the ground. That will result in both M2 and M3 remaining in the triode region. However, we know that when M3 is in deep triode region, its drain-source voltage is not completely zero so it can lead a higher $V_{TH}$ for $M_{Acc2}$. As a result based on Eq.3, discharge rate of the access transistor is slowed down. Therefore, this assumption is not 100% accurate because there is a slight potential difference between the source and bulk of the transistors. When the bulk voltage of an NMOS drops below the source voltage, so long as the source and drain connections remain in a reverse-biased mode, the device continues to operate appropriately. However, there will be some modifications to its characteristics. To comprehend the modifications, first, let us suppose $V_S = V_D = 0$, and $V_G$ is somewhat less than $V_{TH}$ so that a depletion region is formed under the gate, but no inversion layer exists. As $V_B$ decreases towards negative potential, more holes are lured to the substrate connection. Thus, a more significant negative bias will be gathered under the depletion region. A the threshold voltage is a function of the total charge in the depletion region (according to [20]), and as the gate charge must reflect $Q_d$ before an inversion layer is built. Therefore, as $V_B$ drops and $Q_d$ increases, the $V_{TH}$ will also increase. This phenomenon is called the "body effect" or the "back-gate effect,". Thus, the $V_{TH}$ could be achieved based on the following Eq. (6) from [20], that $\gamma = \sqrt{2q\epsilon_{si}N_{sub}}/C_{OX}$ and called body-effect coefficient.

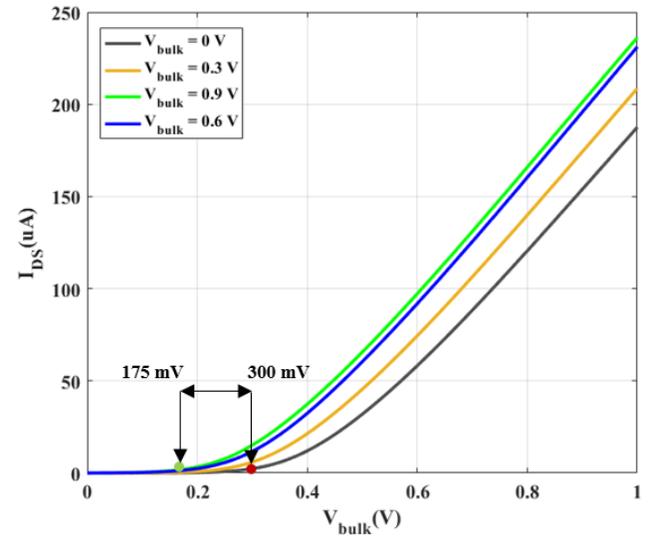

Fig. 3 Body biasing of the access transistor for different $V_{bulk}$

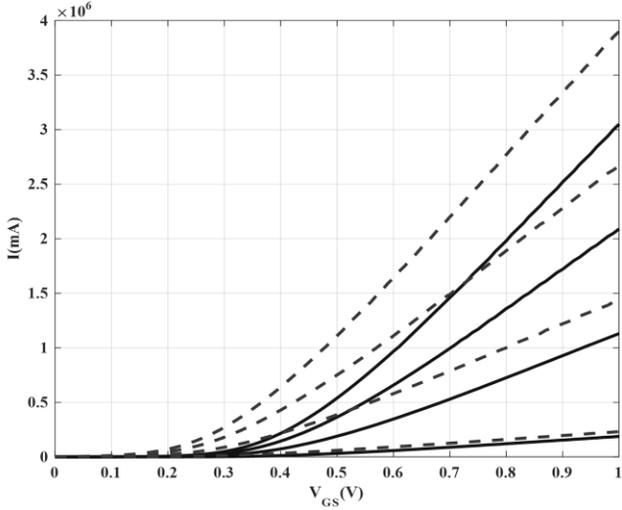

Fig. 4 The sweeping of the transistors' width with $V_{bulk} = 0$ (solid) and $V_{bulk} = 0.6$ V (dashed).

$$V_{TH} = V_{TH0} + \gamma\left(\sqrt{2\phi_F + V_{SB}} - \sqrt{|2\phi_F|}\right) \quad (6)$$

Therefore, to decrease the $V_{TH}$, the only alternative available is to find a way to make $(\sqrt{2\phi_F + V_{SB}} - \sqrt{|2\phi_F|})$ negative. That means the $2\phi_F = -V_{SB}$. It has been depicted in Fig. 3 that by decreasing $V_{TH}$, we facilitate current passing through the cell to start faster, which means that by exploiting body biasing, we could achieve a 125-mV decrease (for 0.6 V body voltage) in the $V_{TH}$ as we intended (see Fig. 3). In Fig. 4, you can see the plot for sweeping the width of the access transistor. As depicted for the case of $V_{Bulk} = 0.6$ (i.e., low $V_{TH}$), the current passing through the cell has been increased (red dashed line) regardless of the transistors' width.

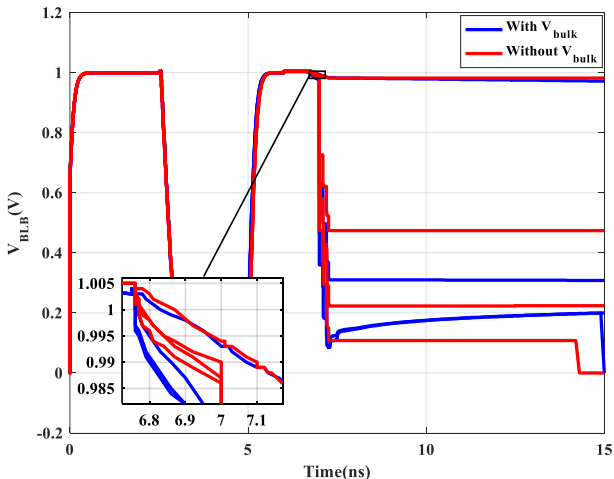

Fig. 6 Body biasing effect on discharge of $V_{BLB}$ for [10]

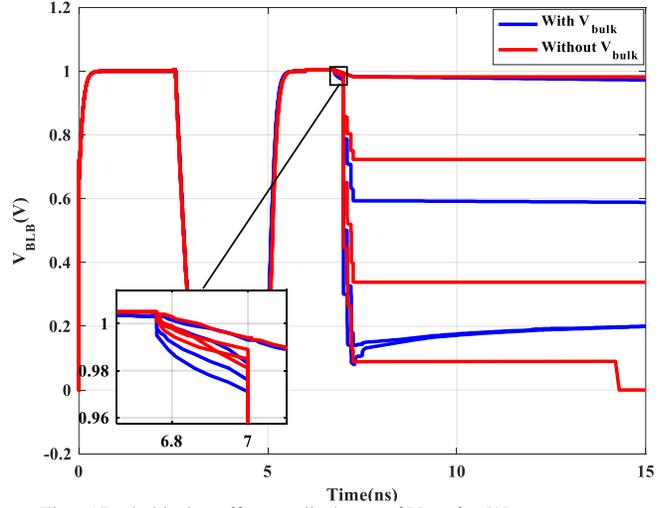

Fig. 5 Body biasing effect on discharge of $V_{BLB}$ for [9]

*C. Investigation of Body effect MAC operation for the State-of-the-art in-SRAM MAC.*

To convey the effect of $V_{TH}$ modification on the result of the MAC operation, firstly, we will explain the process of the operation. One of the operands will be stored first in the SRAM cell. The other operand will be passed through the Digital to Analog Converter (DAC) to produce an analog interpretation of the digital data. State-of-the-art approaches [9], [10] exploit different DAC implementations; however, the available voltage for each input number can be calculated as Eq. 7 and Eq. 8 to operate in the saturation region consecutively.

$$V_{WL}[9] = V_{TH} + V_{DAC} \times \frac{VDD - V_{TH}}{2^N - 1} \quad (7)$$

$$V_{WL}[10] = V_{TH} + \sqrt{V_{DAC} \times \frac{VDD - V_{TH}}{2^N - 1}} \quad (8)$$

In both Eq. (7) and (8), N is the bit-width of the input data;

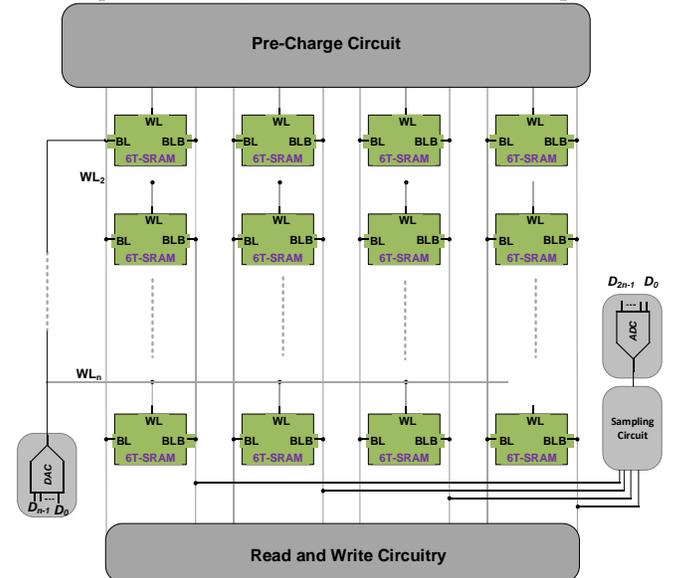

Fig. 7 The architecture of SMART, the body biasing is applied in the access transistor inside the 6T-SRAM structure

therefore, in the case of a four-bit number, it would be equal to fifteen absolute consecutive linear cases. Fig. 6 and Fig. 5 are the simulation results of the most recent state-of-the-art approaches [9], [10]. As seen in both approaches, body biasing caused the discharge of the $V_{BLB}$ to be accelerated. It means that the latency of the operation has been decreased for the same operation. In the next section, we discuss the proposed system architecture of our SMART four by four-bit multiplier. Then we introduce a comparison metric to calculate the BER of architecture.

### III. THE ARCHITECTURE OF SMART

We kept the basic fundamental block of our architecture as a 6T-SRAM, which will be responsible either for a regular read/write operation when the configuration is a memory mode or do a one by-one-bit multiplication when it operates in a mathematical. If the user tries to do multiplication, the write circuit will store one of the operands in the memory cells. Then the memory will switch to the mathematical mode, and the second operand will be passed to the DAC to be coded into an analog representation of a $V_{WL}$. Since the available WL margin has been increased in our technique (i.e., from [300-700] mV in state-of-the-art to [175-700] mV in SMART). The additional achieved margin could either be used for an additional bit width or result in a BER reduction. To facilitate the comparison with [9],[10], we kept the bit-width as a four-by-four-bit multiplication and discussed the BER reduction in the next section. Fig. 7 shows the architecture of our approach, and the green highlighted part is where our novel contribution stands. The gate of the access transistors (i.e., M1acc and M2acc in Fig. 2) has been connected to 0.6 V. It has been provided by exploiting the dual-VDD design technique in the circuit. In this design, the Most Significant Bits (MSBs) are stored in the most left cell, and thus the Least Significant Bits (LSBs) will be stored in the cells located on the right of the architecture. We exploit the designed circuitry of the [10]; therefore, no static current will be applied by the pre-charge circuit, which results in no additional power overhead. Although the timing of the WL has been selected the same as [9], [10] for a fair comparison, the authors believe there is a potential for further improvement considering the optimization of the WL pulse.

### IV. EXPERIMENTAL SETUP, RESULTS, AND COMPARISON WITH THE STATE-OF-THE-ART APPROACHES

The Introduces architecture in the previous section has been implemented via the Cadence Virtuoso 6.1.8 by exploiting the 65nm CMOS technology. The power supply has been configured as 1 V. The gate voltages are selected as 0.6 V. We optimized the circuit parameters, including the sizing of the transistors, $V_{WL}$ via SPECTRE ADE-XL transient simulation considering the Eq. (1) - (8). The optimization was aimed at

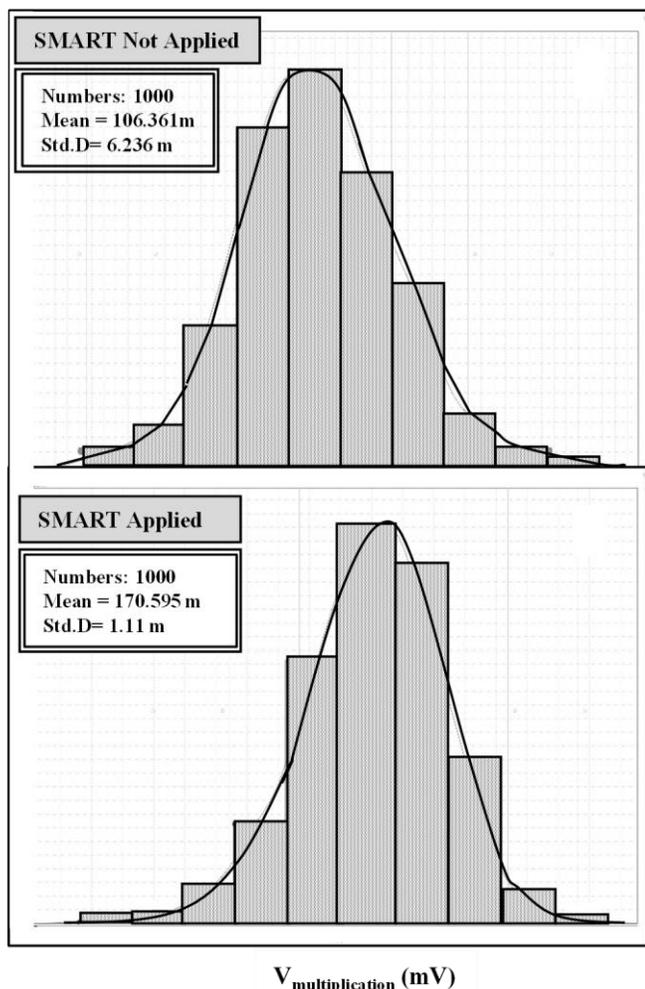

Fig. 8 Accuracy improvement for 1111 * 1111 in [10] exploiting SMART approach

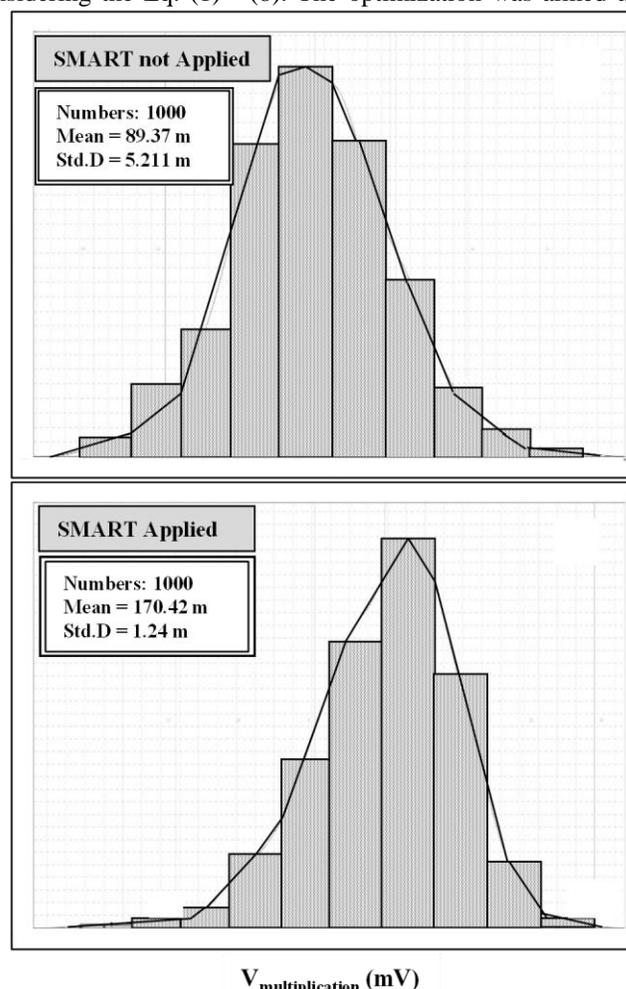

Fig. 9 Accuracy improvement for 1111 * 1111 in [9] exploiting SMART approach

MAC latency improvement, power of the circuit, and the cell area. Notably, there is no area overhead compared with the state-of-the-art circuitry. Since the physical design parameters of the circuit (e.g., $V_{TH}$, $C_{OX}$, and mobility of the electrons) could affect the output result, a 1000 points Monte-Carlo simulation (considering the process and mismatch) has been conducted on the circuit, which the result has been provided in Fig. 8 and Fig. 9. Table 1 compares SMART and state-of-the-art methods for energy per computation, frequency, and accuracy. Accuracy has been calculated based on the proposed SNR in [10]. As can be seen, the accuracy and frequency have been improved.

Table. 1. COMPREHENSIVE COMPARISON OF SMART AND STATEOF-THE-ART TECHNIQUES

|  | SMART | [10] | [9] | [14] | [21] |
|---|---|---|---|---|---|
| Tech. (nm) | 65 | 65 | 65 | 65 | 65 |
| Supply (V) | 1 | 1 | 1.2 | 1 | 1.2 |
| MAC energy (pJ) | 0.783 | 0.523 | 0.9 | 1.3 | 3.5 |
| Accuracy (STD.V) | 0.009 | 0.086 | 0.6 | / | / |
| Frequency (MHz) | 250 | 200 | 100 | 60-125 | 2.5 |

## Conclusion

This paper investigates the effect of threshold voltage suppression on the gate of the access transistors of in-SRAM MAC accelerators. We observed that by exploiting the dual VDD technique to decrease the threshold voltage, an additional read margin for the VWL is achievable. This additional read margin could be used for bit-width increasing or BER improvement. We implemented our proposed architecture at the circuit level exploiting a 65nm CMOS PDK. By performing an intensive Monte-Carlo simulation, including the process and mismatch, we confirmed the efficiency of our proposed method for the state-of-the-art approaches.